\documentclass[amssymb,prd,aps,twocolumn,amsfonts,eqsecnum,nofootinbib]{revtex4}

\usepackage{graphicx}
\usepackage{amsmath}                  
\usepackage{color}
\usepackage{graphics,epsfig}

\newcommand {\dttaa} {{\bar\theta}^{\dot \alpha}}
\newcommand {\dttab} {{\bar\theta}^{\dot \beta}}

\newcommand {\ttaa} {\theta^\alpha}
\newcommand {\ttab} {\theta^\beta}
\newcommand {\ttag} {\theta^\gamma}
\newcommand {\tta} {\theta}
\newcommand {\btta} {\bar \theta}

\newcommand {\smaa} {\sigma^\mu_{\alpha \dot\alpha}}
\newcommand {\smbb} {\sigma^\mu_{\beta \dot\beta}}

\newcommand {\snbb} {\sigma^\nu_{\beta \dot\beta}}

\newcommand {\cab}  {C^{\alpha\beta}}
\newcommand {\cgd}  {C^{\gamma\delta}}
\newcommand {\bcab} {\bar{C}^{\dot\alpha\dot\beta}}
\newcommand {\bcgd} {\bar{C}^{\dot\gamma\dot\delta}}
\newcommand {\al} {\alpha}
\newcommand {\bt} {\beta}
\newcommand {\vt} {\bigg\vert}
\newcommand {\be} {\begin{equation}}
\newcommand {\bear} {\begin{eqnarray}}
\newcommand {\ee} {\end{equation}}
\newcommand {\eear} {\end{eqnarray}}
\newcommand {\besp} {\begin{equation}\begin{split}}
\newcommand {\eesp} {\end{split}\end{equation}}

\begin{document}
%
\preprint{WM-04-117}
\title{Field Theory in Noncommutative Minkowski Superspace
\vskip 0.1in}
\author{Vahagn Nazaryan}\email[]{vrnaza@wm.edu}
\author{Carl E. Carlson}\email[]{carlson@physics.wm.edu}
\affiliation{Particle Theory Group, Department of Physics,
College of William and Mary, Williamsburg, VA 23187-8795}
\date{October 2004}
\begin{abstract}
There is much discussion of scenarios where the space-time
coordinates $x^\mu$ are noncommutative. The discussion has been
extended to include nontrivial anticommutation relations among spinor
coordinates in superspace. A number of authors have studied field
theoretical consequences of the deformation of $\mathcal{N}=1$
superspace arising from nonanticommutativity of coordinates $\tta$,
while leaving $\btta$'s anticommuting. This is possible in Euclidean
superspace only. In this note we present a way to extend the discussion
by making both $\tta$ and $\btta$ coordinates non-anticommuting in
Minkowski superspace.  We present a consistent algebra for the
supercoordinates, find a star-product, and give the
Wess-Zumino Lagrangian $\mathcal{L}_{WZ}$ within our model.  It has two
extra terms due to non(anti)commutativity.  The Lagrangian in Minkowski
superspace is always manifestly Hermitian and for
$\mathcal{L}_{WZ}$ it preserves Lorentz invariance.
\end{abstract}
\maketitle

\section{An Overview and Introduction}\label{sec:intro}

By now, there is a long history of theoretical studies related to
nontrivial, possibly richer structures of spacetime.  Under this
heading one may include supersymmetry and extra-dimensional theories,
but we concentrate here on theories with a noncommutative spacetime algebra.  The earliest motivation for such theories was the hope that divergences in field theory would be be ameliorated if there were coordinate uncertainty, and coordinate uncertainty would follow if coordinate operators did not commute~\cite{snyder}.  The idea did not bear direct fruit, and Snyder's paper~\cite{snyder} remained almost alone for many decades.

Recently, the idea of noncommutative coordinates has blossomed, at least as theoretical speculation, with motivation from several sources.  For example, Connes {\it et al.}~\cite{Connes} attempted to make gauge theories of electroweak unification mathematically more natural by using ideas from noncommutative geometry.  Also, Dopplicher, Fredenhagen, and Roberts~\cite{dop} saw general relativity as giving a natural limit to the precision of locating a particle, which to them suggested an uncertainty relation and noncommutativity among coordinate operators.  They suggested a particular algebra of the coordinates now often referred to as the ``DFR'' algebra.  However, probably the greatest modern spur to studying spacetime noncommutativity was the observation that string theories in a background field can be solved exactly and give coordinate operators which do not commute~\cite{ardalan,SW}.

In theories with an underlying noncommutative spacetime algebra, the position
four vector $x^\mu$ is promoted to an operator $\hat{x}^\mu$ that satisfies the commutation relation
\be
[\hat{x}^\mu,\hat{x}^\nu]= \Theta^{\mu\nu} \label{canonical-NC}.
\ee
The $\Theta^{\mu\nu}$ that comes out of string theory, which is directly related 
to the background field $B^{\mu\nu}$~\cite{SW}, is just an antisymmetric array of c-numbers.  There has been a fair amount of theoretical study learning how to work with fields that are functions of noncommuting coordinates, and phenomenological studies of possible physical consequences of spacetime noncommutativity.   However, theories based 
on~(\ref{canonical-NC}) with a c-number $\Theta^{\mu\nu}$ suffer from 
Lorentz-violating effects.  Such effects  are severely 
constrained~\cite{MPR}\nocite{Chaichian,HK,CHK,CCL,ABDG,Carlson:2002zb,VN,h2,Bertolami:2003nm}-\cite{Carone:2004wt}
by a variety of low energy experiments~\cite{LeExp}.

Returning to one of our previous remarks, in the DFR noncommutative
algebra~\cite{dop} $\hat{x}^\mu$ satisfies $[\hat{x}^\mu,\hat{x}^\nu]=
\hat{\Theta}^{\mu\nu}$, but where here
$\hat{\Theta}^{\mu\nu}=-\hat{\Theta}^{\nu\mu}$ transforms as a Lorentz
tensor and is in the same algebra with $\hat{x}^\mu$. Thus the algebra
formulated by DFR is Lorentz-invariant.   Carone, Zobin, and one of the
present authors (CEC)~\cite{CCZ} formulated and studied some
phenomenological consequences of a Lorentz-conserving noncommutative
QED (NCQED) based on a contracted Snyder~\cite{snyder} algebra, which
has the same Lie algebra as DFR. In~\cite{CCZ} light-by-light
scattering was studied, and it was found that contributions from
noncommutativity can be significant with respect to the standard model
background. Further studies of NCQED as formulated in~\cite{CCZ} may be found
in~\cite{morita,CKN,Haghighat}. In particular, bounds were obtained on
the scale of noncommutativity~\cite{CKN} in the Lorentz conserving case
from an number of QED processes for which there exist experiments at the CERN Large Electron and Positron collider (LEP).

There have also been studies extending noncommutativity to the full set of supersymmetric coordinates, not just limiting noncommutativity to ordinary spacetime.   In this paper, we wish to continue the study of noncommutative coordinates in supersymmetric theories, by giving and studying consequences of an algebra of superspace coordinates that very definitely allows us to remain in Minkowski space.

Recent work (e.g.,~\cite{OoVa, deBoer, Seiberg, Berkovits, Imaanpur}) has stimulated interest in supersymmetric noncommutativity by showing, in Euclidean space, how noncommutative supercoordinates could arise from string theory.   Further, some of the recent work~\cite{Seiberg} defined a star-product from the commutation relations.  Operators multiplied in noncommutative space could then be replaced by their symbols in commutative space with multiplication replaced by the star-product.  This was then used to study noncommutative modifications to Wess-Zumino and gauge Lagrangians, albeit still in Euclidean space.  Proofs of renormalizability of the deformed Wess-Zumino Lagrangian were offered~\cite{Berenstein}, but it was noted that the deformed Euclidean space Lagrangians, as well as the vector superfield, were not Hermitian~\cite{Berenstein,Araki}.

Working in Euclidean space allows coordinates $\theta$ with nontrivial anticommutators to be paired with $\bar\theta$'s that anticommute in the normal way; the phrase $N=1/2$ supersymmetry described this.  There is no direct analog in Minkowski space, where the $\theta$'s and $\bar\theta$'s are tightly connected.  

Useful formal developments include, using the star-product to define the theory, a display of a number of different ways to introduce noncommutativity into superspace~\cite{Ferrara,Park:2003ku,Chaichian:2003dp}.   Also~\cite{Klemm} showed that in Minkowski space nontrivial anticommutation relations for the $\theta$'s and $\bar\theta$'s were not compatible with having an associative algebra.  Hence we have some freedom in the choice of a star-product, but must be open to using a star-product that is non-associative.

In the next section, Sec.~\ref{sec:algebra}, we present a consistent
set of (anti)commutation relations among the supercoordinates in
Minkowski space.   Following that, Sec.~\ref{sec:star} defines our
theory by presenting a star product that yields the deformed
supercoordinate algebra developed in section~\ref{sec:algebra}.   We
record the deformed algebra of supersymmetry generators, and of the
covariant superderivatives.   The commutators of the supergenerators
and superderivatives break supersymmetry.  In Sec.~\ref{sec:WZ} we
write down the chiral and antichiral superfields, and show that
products of (anti)chiral superfields are themselves (anti)chiral
superfields.  This is a feature retained from commutative
supersymmetry; some of the choices in Sec.~\ref{sec:algebra} were in
fact made in the hope that this would happen.   We construct the
Wess-Zumino Lagrangian $\mathcal{L}_{WZ}$, and show how to avoid
ambiguity in our construction despite the nonassociativity of the
products.   We end with some  discussion in section~\ref{sec:summary}.

\section{The Non(anti)commutative SUSY Algebra}\label{sec:algebra}					   %

Noncommutativity has usually been studied as the noncommutativity of ordinary spacetime.  Here we are considering noncommutativity in superspace\footnote{We follow conventions of Wess and Bagger~\cite{Wess-Bagger}.}, and for Minkowski rather than Euclidean space.   The supercoordinate is $(x^\mu, \theta^\alpha, \bar \theta^{\dot\alpha})$ where $\theta^\alpha$ and $\bar\theta^{\dot\alpha}$ are normally anticommuting Grassmann variables that we shall promote to nonanticommuting operators $\hat\theta^\alpha$ and $\hat{\bar\theta}^{\dot\alpha}$ in some algebra.

The anticommutation for the $\hat\tta$'s will be
\be
\{\hat\tta^\alpha,\hat\tta^\beta\}=C^{\alpha\beta}, \label{com-tta}
\ee
%
where $C^{\alpha\beta}$ is a symmetric array of c-numbers.  We shall also suppose there is a mapping between the operator $\hat\theta^\alpha$ and a Grassmann variable $\theta^\alpha$ in ordinary (anti)commutative space.  We will soon, as usual, obtain using commutative variables the multiplication rules of the noncommutative algebra by using a star-product rather than the ordinary product for variables and functions in commutative space.

In Minkowski space, we relate $\hat{\bar\theta}^{\dot\alpha}$ to $\hat\theta^\alpha$ by
\be
\hat{\bar\theta}^{\dot\alpha} = (\hat\theta^{\alpha})^\dagger \ ,
\ee 
so that the $\hat{\bar\theta}^{\dot\alpha}$ are noncommutative also,
\be
\{ \hat{\bar\theta}^{\dot \alpha} ,\hat{\bar\theta}^{\dot \beta} \}=
				       \bar C^{\dot\alpha \dot\beta},
\label{com-btta}
\ee
where $\bar{C}^{\dot\alpha \dot\beta}=(C^{\beta\alpha})^*$.

The commutators of $\hat\theta$ and $\hat{\bar\theta}$ are still unconstrained, and we make the simple choice
\be
\{ \hat{\bar{\theta}}^{\dot{\alpha}}, \hat\tta^\alpha \}=0.
\label{com-tta-btta}
\ee

Next we fix the commutation relations among $\tta$'s and 
spacetime coordinates. We define the commutator of the chiral coordinate 
$\hat{y}^\mu \equiv \hat{x}^\mu+i\hat{\tta}\sigma^\mu\hat{\btta}$ with $\hat{\tta}$, and the commutator of the antichiral coordinate  
$\hat{\bar{y}}^\mu \equiv \hat{x}^\mu-i\hat{\tta}\sigma^\mu\hat{\btta}$ with $\hat{\bar{\tta}}$, in such a way that enables us to write products of chiral fields, and products of antichiral fields, in their canonical form.  We choose
\begin{align} 
[\hat{y}^\mu,\hat{\tta}^\al]&=0, \label{yhat} \\
[\hat{\bar{y}}^\mu,\hat{\btta}^{\dot\al}]& =0\, .  \label{ybarhat}
\end{align}
%
The nonzero commutators
\be \label{ybarhat-tta}
 [\hat{\bar{y}}^\mu,\hat{\tta}^\al]= -2[i\hat{\tta}\sigma^\mu \hat{\btta},
                                                                         \hat{\tta}^\al] 
			    = 2iC^{\al\bt}\smbb\hat{\btta}^{\dot\bt},
\ee
and 
\be \label{yhat-btta}
[\hat{y}^\mu,\hat{\btta}^{\dot\al}]= 2[i\hat{\tta}\sigma^\mu \hat{\btta},
                                                                         \hat{\btta}^{\dot\al}]
			      =2i\bar{C}^{\dot\al\dot\bt} \tta^\bt\smbb \ ,
\ee
are fixed by the choices already made.

The choices and results in~(\ref{com-tta})-(\ref{ybarhat}) also constrain the 
commutation relations of $\hat{y}$ and of $\hat{\bar{y}}$ with themselves.  
The following condition must be satisfied:
\be \label{y-ybar}
[\hat{y}^\mu,\hat{y}^\nu] - [\hat{\bar{y}}^\mu,\hat{\bar{y}}^\nu]
=4(\bcab\hat{\tta}^\al \hat{\tta}^\bt
      - \cab  \hat{\btta}^{\dot\al} \hat{\btta}^{\dot\bt})\smaa \snbb \ .
\ee

\noindent
Thus, the Hermitian part of $[\hat{y}^\mu,\hat{y}^\nu]$ is fixed by choices already made.  
Let us rewrite the previous equation in the following way,
\begin{equation} \label{y-ybar-2}
\begin {split}
[\hat{y}^\mu,\hat{y}^\nu] - [\hat{\bar{y}}^\mu,\hat{\bar{y}}^\nu]
& =(4\bcab\hat{\tta}^\al\hat{\tta}^{\bt}-2\cab\bcab) \smaa\snbb\\
& \; +(4\cab\hat{\btta}^{\dot{\bt}} \hat{\btta}^{\dot\al}-2\cab\bcab)\smaa\snbb \ ,
\end{split}
\end{equation}
%
where each term on the right-hand-side is the Hermitian conjugate of the other.
Then we make the choices, 
\be \label{yy}
[\hat{y}^\mu,\hat{y}^\nu] = (4\bcab\hat{\tta}^\al\hat{\tta}^{\bt}-2\cab\bcab)\smaa\snbb \ ,
\ee
and
\be \label{ybarybar}
[\hat{\bar{y}}^\mu,\hat{\bar{y}}^\nu] = (4\cab\hat{\btta}^{\dot{\al}} \hat{\btta}^{\dot\bt}
                                                                             -2\cab\bcab)\smaa\snbb \ ,
\ee
which are natural and consistent with already defined commutators.  Finally, note that $\hat{y}$ and $\hat{\bar{y}}$ do not commute in this non(anti)commutative algebra,
\be \label{yybar}
[\hat{y}^\mu,\hat{\bar{y}}^\nu] = 2\cab\bcab\smaa\snbb \ ,
\ee
although their commutator is a c-number.

Commutation relations given by~(\ref{com-tta})-(\ref{yhat-btta}), (\ref{yy}) and (\ref{ybarybar}) are compete, consistent with each other, and represent the deformed supersymmetry algebra in terms of chiral and spinor variables. One can summarize this algebra in terms of $(\hat{x},\hat{\tta},\hat{\btta})$ as, 
\begin{align} 
\{\hat\tta^\alpha,\hat\tta^\beta\}& =C^{\alpha\beta} \,, \qquad 
[\hat{x}^\mu, \hat{\tta}^\al] = i\cab\smbb\hat{\btta}^{\dot{\bt}}\,, \label{xtta} \\
\{ \hat{\bar\theta}^{\dot \alpha} ,\hat{\bar\theta}^{\dot \beta} \}&= \bcab \,, \qquad
[\hat{x}^\mu, \hat{\btta}^{\dot{\al}}] = i\bcab\hat{\tta}^\bt \smbb \,, \label{xbtta} \\
\{ \hat{\bar{\theta}}^{\dot{\alpha}}, \hat\tta^\alpha \}&=0 \,,  \nonumber \\
[\hat{x}^\mu,\hat{x}^\nu] &=  
		(\cab \hat{\btta}^{\dot\al} \hat{\btta}^{\dot\bt} - 
                                                            \bcab \hat{\tta}^\bt \hat{\tta}^\al)\smaa\snbb\,. \label{xx}
\end{align}


Hence, the space-time coordinates $x^\mu$ do not commute with each other either, or with the spinor coordinates $\tta$ and $\btta$. 


\section{The Star Product} \label{sec:star}					   		   %


We shall assume that there exists a mapping that relates the ordinary variables $({x},{\tta},{\btta})$in commutative to their counterparts $(\hat{x},\hat{\tta},\hat{\btta})$ in noncommutative space, and that relates functions $f(x,\tta,\btta)$ in commutative space to operators $\hat{f}(\hat{x}, \hat{\tta}, \hat{\btta})$ in the noncommutative algebra.  Products of functions in commutative space will be defined by a star-product.   In noncommutative theories a star product is used so that the result of products such as
$\hat{f}(\hat{x}, \hat{\tta}, \hat{\btta})\,\hat{g}(\hat{x}, \hat{\tta}, \hat{\btta})\,
\hat{h}(\hat{x},\hat{\tta}, \hat{\btta})$ 
in noncommutative space corresponds to the result of
$f(x,\tta,\btta)*g(x,\tta,\btta)*h(x,\tta,\btta)$ in commutative space (provided $\hat{f}(\hat{x}, \hat{\tta}, \hat{\btta})$ corresponds to $f(x,\tta,\btta)$, {\it etc.}). 

We operationally define our theory by finding a suitable star-product.
A properly defined star product has to reproduce the underlying
deformed algebra of the supercoordinates in its
entirety.  We will require that the star product satisfy the
reality condition, that is, the star-product will maintain the usual
rules for products of involutions,
\be \label{reality}
(f_1 * f_2)^\dagger = f_2^\dagger * f_1^\dagger \ .
\ee
We find it convenient to use the supersymmetry generators in defining the star product, and will limit the star-product to being at most quadratic in deformation parameter $C$.  This is also the minimum that will allow reproducing the deformed algebra for the supercoordinates.

Before we define the star product, we find it useful to have before us the well known
canonical expressions for covariant derivatives and supercharges.  Acting on the right,
\bear
\roarrow{D}_\alpha &=& \frac{\roarrow{\partial}}{\partial\tta^\alpha}\vt_x+
		    i\smaa\dttaa
		    \frac{\roarrow{\partial}}{\partial x^\mu}, \nonumber \\
\roarrow{\bar{D}}_{\dot\alpha} &=& -\frac{\roarrow{\partial}}{\partial\dttaa}\vt_x-
		    i\tta^\alpha \smaa
		    \frac{\roarrow{\partial}}{\partial x^\mu},
\label{Dx}
\eear
and
\bear
\roarrow Q_\alpha &=& \frac{\roarrow{\partial}}{\partial\tta^\alpha}\vt_x -
		    i\smaa\dttaa
		    \frac{\roarrow{\partial}}{\partial x^\mu}, \nonumber \\
\roarrow{\bar{Q}}_{\dot\alpha} &=& -\frac{\roarrow{\partial}}{\partial\dttaa}\vt_x +
		    i\tta^\alpha \smaa
		    \frac{\roarrow{\partial}}{\partial x^\mu}.
\label{Qx}
\eear
In (\ref{Dx}) and (\ref{Qx}) derivatives with respect to $\tta$ and 
$\btta$ are taken at fixed $x$, and derivatives with respect to $x$ are taken 
at fixed $\tta$ and $\btta$.

Let's also write down the corresponding equation for two sets of coordinates 
$(y,\ttaa,\dttaa)$ and $(\bar y,\ttaa,\dttaa)$, where 
\be
y^\mu = x^\mu + i\tta\sigma^\mu \btta,  \quad
{\bar y}^\mu = x^\mu - i\tta\sigma^\mu \btta .
\ee
Then one can check that
\begin{align}
\roarrow{D}_\alpha &= \frac{\roarrow\partial}{\partial \tta^\alpha } \vt_y + 
		     2i\smaa\dttaa
		     \frac{\roarrow\partial}{\partial y^\mu}\,\, , \quad
\roarrow D_\alpha = \frac{\roarrow\partial}{\partial\tta^\alpha} \vt_{\bar y}\,\, , \label{D-y-ybar}\\
\roarrow{\bar{D}}_{\dot\alpha} &= -\frac{\roarrow\partial}{\partial\dttaa}
				   \vt_y\,\, , \quad
\roarrow{\bar{D}}_{\dot\alpha} = -\frac{\roarrow\partial}{\partial\dttaa}\vt_{\bar y}
				  -2i\ttaa \smaa
				  \frac{\roarrow\partial}{\partial {\bar y}^\mu}\,\, , \label{Dbar-y-ybar}\\
\roarrow{Q}_\alpha & =\frac{\roarrow\partial}{\partial \ttaa } \vt_y\,\, , \quad
\roarrow{Q}_{\alpha}  =\frac{\roarrow\partial}{\partial\ttaa}\vt_{\bar y}
				  -2i\smaa\dttaa
				  \frac{\roarrow\partial}{\partial {\bar y}^\mu}\,\, , \label{Q-y-ybar}\\
\roarrow{\bar{Q}}_{\dot\alpha} &= -\frac{\roarrow\partial}{\partial\dttaa}\vt_y
				  +2i\ttaa \smaa
				  \frac{\roarrow\partial}{\partial y^\mu}\,\, , \quad
\roarrow{\bar{Q}}_{\dot\alpha} =-\frac{\roarrow\partial}{\partial\dttaa}\vt_{\bar y}\,\, .
\label{Qbar-y-ybar}
\end{align}

Expressions for $\loarrow{D}_\alpha,\loarrow{\bar{D}}_{\dot\alpha},
\loarrow{Q}_\alpha,\text{ and }\loarrow{\bar{Q}}_{\dot\alpha}$ are obtained from 
above by simply changing $\rightarrow$ to $\leftarrow$, with the following definitions,
\begin{align}
\frac{\roarrow\partial}{\partial \ttaa }\ttab &\equiv \delta^{\beta}_{\alpha}\,\, ,&
\ttab\frac{\loarrow\partial}{\partial \ttaa } &\equiv -\delta^{\beta}_{\alpha}\,\, ,\\
\frac{\roarrow\partial}{\partial y^\mu }y^\nu &\equiv \delta^\nu_\mu\,\, ,&
y^\nu\frac{\loarrow\partial}{\partial y^\mu } &\equiv \delta^\nu_\mu\,\, .
\end{align}
Similar definitions apply derivatives with respect to $\dttaa$ and $\bar{y}^\mu$.

Now we can write down the star product that we use for mapping a product
of functions $\hat{f}\hat{g}$ in noncommutative space to a product of functions in commutative space.
\be \label{star}
\hat{f} \hat{g} \Rrightarrow f * g = f  \left( 1 + \cal{S}
		\right)g \, . 
\ee
Here $f$ and $g$ can be functions of any of the three sets of variables mentioned above,
and the extra operator $\cal S$ is 
\begin{equation} \label{exponent}
\begin{split}
{\cal S} &=
   		-\frac{C^{\alpha\beta}}{2}
		{\loarrow Q_\alpha}\roarrow{Q_\beta} 
		-\frac{\bar C^{\dot\alpha\dot\beta}}{2}
		\loarrow{\bar{Q}}_{\dot\alpha}\roarrow{\bar{Q}}_{\dot\beta} \\
& \quad  +\frac{\cab \cgd}{8} \loarrow{Q}_\alpha \loarrow{Q}_\gamma   
                                                     \roarrow{Q}_\delta  \roarrow{Q}_\beta 
+ \frac{\bcab \bcgd}{8} \loarrow{\bar{Q}}_{\dot\alpha} \loarrow{\bar{Q}}_{\dot\gamma}
		     \roarrow{\bar{Q}}_{\dot\delta} \roarrow{\bar{Q}}_{\dot\beta} \\
& \quad + \frac{\cab \bcab}{4}\left( \loarrow{\bar{Q}}_{\dot\alpha}\loarrow{Q}_\alpha
			       \roarrow{\bar{Q}}_{\dot\beta} \roarrow{Q}_\beta 
			       -\loarrow{Q}_\alpha \loarrow{\bar{Q}}_{\dot\alpha}
			        \roarrow{Q}_\beta  \roarrow{\bar{Q}}_{\dot\beta} \right)  .
\end{split}
\end{equation} 
%
It is easy to verify that the star product presented above indeed reproduces the entire
noncommutative algebra of supersymmetry parameters, and that it satisfies the reality 
condition~(\ref{reality}).

 If $f$ and $g$ are functions only of $\tta$ or only of $\btta$, then the star product takes the following simple forms, recognizable from~\cite{Seiberg},
\be
\begin{split}
f(\tta)\,*\,g(\tta)&=
	f(\tta)\bigg(1-\frac{C^{\al\bt}}{2}\frac{\loarrow\partial}{\partial \ttaa }
		   \frac{\roarrow\partial}{\partial \ttab }
		   													\\  
		 & \quad -\det C\frac{\loarrow\partial}{\partial \tta\tta }
			      \frac{\roarrow\partial}{\partial \tta\tta }\bigg)g(\tta)
	  														\\
&= f(\tta) \exp
	\left(-\frac{C^{\al\bt}}{2}\frac{\loarrow\partial}{\partial \ttaa }
	  \frac{\roarrow\partial}{\partial \ttab }\right)g(\tta) \,,
\end{split}
\ee
and
\be
\begin{split}
f(\btta)\,*\,g(\btta)&=
	f(\btta)\bigg(1-\frac{{\bar 	C}^{\dot\al\dot\bt}}{2}
		\frac{\loarrow\partial}{\partial \dttaa }
		   \frac{\roarrow\partial}{\partial \dttab }		\\
	& \quad
		-\det {\bar C}\frac{\loarrow\partial}{\partial \btta\btta }
	\frac{\roarrow\partial}{\partial \btta\btta }\bigg)g(\btta)
	  															\\
&= f(\btta) \exp \left(-\frac{{\bar C}^{\dot\al\dot\bt}}{2}\frac{\loarrow\partial}
{\partial \dttaa }
	  \frac{\roarrow\partial}{\partial \dttab }\right)g(\btta)
	  \,,
\end{split}
\ee
where
\be
\frac{\partial}{\partial \tta\tta }\equiv \frac{1}{4}\frac{\partial}{\partial\tta_\al}
\frac{\partial}{\partial\ttaa}= \frac{1}{4}\epsilon^{\gamma\eta}
\frac{\partial}{\partial\ttag}\frac{\partial}{\partial \tta^\eta}\,,
\ee
and
\be
\frac{\partial}{\partial \btta\btta }\equiv \frac{1}{4}\frac{\partial}{\partial\dttaa}
\frac{\partial}{\partial\btta_{\dot\al}}= -\frac{1}{4}\epsilon^{\dot\gamma\dot\eta}
\frac{\partial}{\partial\btta^{\dot\gamma}}\frac{\partial}{\partial \btta^{\dot\eta}}\,.
\ee 

The following equations are useful for deriving commutation relations 
among various coordinates of deformed superspace, 

\bear
\ttaa\,*\,\ttab &=& 
-\frac{1}{2}\epsilon^{\alpha\beta}\tta\tta+\frac{1}{2}C^{\alpha\beta}\,, \label{tt}\\
\dttaa\, *\,\dttab &=&
+\frac{1}{2}\epsilon^{\dot\alpha\dot\beta}\btta\btta+\frac{1}{2}{\bar C}^{\dot\alpha\dot\beta}\,. 
\label{tbtb}
\eear

Also,
\be 
\ttaa \,*\, \tta\tta = C^{\alpha\beta}\tta_\beta \,, \qquad
\dttaa \,*\, \btta\btta = -{\bar C}^{\dot\al\dot\beta}\btta_{\dot\beta} \,,
\ee
\be
\tta\tta\,*\,\ttaa = -C^{\alpha\beta}\tta_\beta \,, \qquad
\btta \btta \,*\, \dttaa = {\bar C}^{\dot\al\dot\beta}\btta_{\dot\beta} \,,
\ee

\be
\begin{split}
\tta \tta \,*\,\tta\tta &= -\frac{1}{2}\epsilon_{\al\al^\prime}
			    \epsilon_{\bt\bt^\prime} C^{\al\bt} C^{\al^\prime\bt^\prime} \\
&=- \det C \,,  \\
\btta \btta \,*\,\btta\btta &= -\frac{1}{2}\epsilon_{\dot\al\dot\al^\prime}
			       \epsilon_{\dot\bt\dot\bt^\prime}{\bar C}^{\dot\al\dot\bt}
			       {\bar C}^{\dot\al^\prime\dot\bt^\prime}   \\
&=- \det {\bar C}\,.
\end{split}
\ee
%
and
\be
\begin{split}
\tta\sigma^\mu\btta\,*\,\tta\sigma^\nu\btta   &=
-\frac{1}{2}\tta\tta\btta\btta\eta^{\mu\nu}-\frac{1}{2}\tta\tta{\bar C^{\mu\nu}}
-\frac{1}{2}\btta\btta C^{\mu\nu}
\\ & \quad 
-\frac{1}{4}C^{\alpha\beta}{\bar C^{\dot\alpha\dot\beta}}
\smaa\snbb\,, \label{tst}
\end{split}
\ee
where $C^{\mu\nu}$ and ${\bar C}^{\mu\nu}$ are defined as
\be
C^{\mu\nu} \equiv \frac{1}{4}C^{\alpha\beta}\epsilon_{\beta\gamma}
\left(\sigma^\mu\bar{\sigma}^\nu-\sigma^\nu\bar\sigma^\mu\right)_\al^{
\phantom{\al}\gamma}
=C^{\alpha\beta}\epsilon_{\beta\gamma}(\sigma^{\mu\nu})_\al^{
\phantom{\al}\gamma}\,,
\ee
\be
\bar{C}^{\mu\nu} \equiv \frac{1}{4}\bar{C}^{\dot\alpha\dot\beta}
\epsilon_{\dot\beta\dot\gamma}
(\bar\sigma^\mu\sigma^\nu-\bar\sigma^\nu\sigma^\mu)^{\dot\gamma}_{
\phantom{\dot\gamma}\dot\alpha}
=\bar{C}^{\dot\alpha\dot\beta}\epsilon_{\dot\beta\dot\gamma}
(\bar\sigma^{\mu\nu})^{\dot\gamma}_{\phantom{\dot\gamma}\dot\alpha}\,.\label{CC}
\ee

One can now verify,
\begin{align} 
\{\tta^\alpha, \tta^\beta\}_* & =C^{\alpha\beta} \,, &
[x^\mu,  {\tta}^\al]_* &= i\cab\smbb \btta^{\dot{\bt}}\,, \label{xtta-star} \\
\{ {\bar\theta}^{\dot \alpha} , {\bar\theta}^{\dot \beta} \}_* &= \bcab \,, &
[x^\mu, \btta^{\dot{\al}}]_* &= i\bcab \tta^\bt \smbb \,, \label{xbtta-star} \\
\{ {\bar{\theta}}^{\dot{\alpha}}, \tta^\alpha \}_* &=0 \,, &
[x^\mu, x^\nu]_* &=  \btta\btta C^{\mu\nu} + \tta\tta \bar{C}^{\mu\nu} \,. \label{xx-star} 
\end{align}
as they should be according to~(\ref{xtta})-(\ref{xx}).  Subscript ``$*$'' means use star multiplication when evaluating the (anti)commutators.

From~(\ref{Q-y-ybar}), and (\ref{Qbar-y-ybar}) one may check that in noncommutative space
\begin{align} 
\{Q_\al,Q_\bt \}&= -4\bcab \smaa \snbb \frac{\partial^2}{\partial \bar{y}^\mu \partial \bar{y}^\nu} \,,  \label{QQ} \\
\{{\bar Q}_{\dot\al},{\bar Q}_{\dot\bt}\}&= -4\cab \smaa \snbb \frac{\partial^2} {\partial y^\mu \partial y^\nu} \,, 
\label{QbarQbar} \\
\{\roarrow{Q}_\al,\roarrow{\bar{Q}}_{\dot\al}\} &= 2i\smaa \frac{\partial}{\partial y^\mu} \,. \label{QQbar}
\end{align}
Thus, we see that the first two of the above three anticommutators of supercharges are deformed from their 
canonical forms. From~(\ref{D-y-ybar}), and (\ref{Dbar-y-ybar}) for the covariant derivatives we find,
\begin{align}
\{D_\al,D_\bt \}&=0 \,, \label{DD} \\
\{\bar{D}_{\dot\al},\bar{D}_{\dot\bt}\}& =0 \,, \label{DbarDbar} \\
\{\roarrow{D}_\al,\roarrow{\bar{D}}_{\dot\al}\} & = -2i\smaa \frac{\partial}{\partial y^\mu} \,. \label{DDbar}
\end{align}
So, the anticommutators of covariant derivatives are not deformed in this noncommutative superspace.
The anticommutators of supercharges and covariant derivatives with each other
are not deformed either,
\be
\{ D_\al, Q_\bt\}=\{ \bar{D}_{\dot\al}, Q_\bt \}=\{ D_\al, \bar{Q}_{\dot\bt}\}
=\{\bar{D}_{\dot\al},\bar{Q}_{\dot\bt} \}=0\,. \label{DQ}
\ee
Hence, we can still define supersymmetry covariant constraints on
superfields as in commutative supersymmetric theory, using the
following defining equations for chiral and antichiral superfields as
before,
\bear
{\bar D}_{\dot\al}\Phi (y,\tta) &=& 0\,,\label{chiral}\\ 
D_\al {\bar \Phi}({\bar y},\btta)&=&0\,.\label{antichiral}
\eear

\section{The Wess-Zumino Lagrangian} \label{sec:WZ}			   			   %

\subsection{Chiral and Antichiral Superfields}

Chiral $\Phi(\hat{y},\hat{\tta})$ and antichiral $\bar{\Phi}(\hat{\bar{y}},\hat{\bar{\tta}})$ 
superfields satisfy (\ref{chiral}) and (\ref{antichiral}) respectively.  We may expand  $\Phi(\hat{y},\hat{\tta})$ and $\bar{\Phi}(\hat{\bar{y}},\hat{\bar{\tta}})$ as a power series in 
$\hat\tta$ and $\hat\btta$.  Just as in commutative theory, no term in the series will have more than two powers of $\hat\tta$ and $\hat\btta$.  In noncommutative theory, this is true because
products with three or more factors of $\hat\theta$ can be reduced to sums of terms with two or fewer $\hat\theta$, and similarly for $\hat{\bar\theta}$.    
Hence, 
\bear
\Phi(\hat{y},\hat{\tta}) &=& A(\hat{y}) + \sqrt{2}\hat{\tta}\psi(\hat{\tta}) +
			     \hat{\tta}\hat{\tta}F(\hat{y})\,,\label{chiral_s}\\
\bar{\Phi}(\hat{\bar{y}},\hat{\bar{\tta}}) &=& A(\hat{\bar{y}}) + 
				 \sqrt{2}\hat{\btta}\bar{\psi}(\hat{\bar{y}})+
				 \hat{\btta}\hat{\btta}\bar{F}(\hat{\bar{y}})\,.\label{antichiral_s}
\eear 
%
The combination $\hat\tta\hat\tta$ is already Weyl ordered, and maps simply into $\theta\theta$ in commutative space.

From~(\ref{star}), the product of two chiral and the product of two antichiral fields is,

\begin{widetext}

\begin{equation} \label{chiral-1-2}
\begin{split}
\Phi_1(y,\tta)\, * \,\Phi_2(y,\tta)&={\Phi_1}(y,\tta)\Phi_2(y,\tta)
-C^{\al\bt}\psi_{1\al}\psi_{2\bt} -{\textrm{det}}C F_1F_2 \\
& \quad + \sqrt{2}\tta^\gamma\cab\big[\epsilon_{\bt\gamma}(\psi_{1\al}F_2 - \psi_{2\al}F_1)  
  +\bcab \smaa\sigma_{\gamma\dot\bt}^\nu(\partial_\mu A_1 \partial_\nu \psi_{2\bt} 
					 - \partial_\mu A_2 \partial_\nu \psi_{1\bt} ) \big]\\
 & \quad +\tta\tta\big[ 2\bar{C}^{\mu\nu}\partial_\mu A_1 \partial_\nu A_2
+\cab\bcab \smaa\sigma_{\bt\dot\bt}^\nu(\partial_\mu A_1 \partial_\nu F_2 
					 - \partial_\mu A_2 \partial_\nu F_1 )\big] \,,
\end{split}
\end{equation}
and
\begin{equation} \label{antichiral-1-2}
\begin{split}
\bar{\Phi}_1(\bar{y},\btta)\, * \,\bar{\Phi}_2(\bar{y},\btta)&=\bar{\Phi}_1(\bar{y},\btta)\bar{\Phi}_2(\bar{y},\btta)
-\bcab \bar{\psi}_{1\dot\al}\bar{\psi}_{2\dot\bt} -{\textrm{det}}\bar{C} \bar{F}_1\bar{F}_2 \\
& \quad + \sqrt{2}\btta^{\dot\gamma}\bcab\big[\epsilon_{\dot\bt\dot\gamma}
(\bar{\psi}_{1\dot\al}\bar{F}_2 - \bar{\psi}_{2\dot\al}\bar{F}_1)  
  +\cab \smaa\sigma_{\bt\dot\gamma}^\nu(\partial_\mu \bar{A}_1 \partial_\nu \bar{\psi}_{2\dot\bt} 
					 - \partial_\mu \bar{A}_2 \partial_\nu \bar{\psi}_{1\dot\bt} ) \big]\\
& \quad +\btta\btta\big[ 2C^{\mu\nu}\partial_\mu \bar{A}_1 \partial_\nu \bar{A}_2
+\cab\bcab \smaa\sigma_{\bt\dot\bt}^\nu(\partial_\mu \bar{F}_1 \partial_\nu \bar{A}_2 
					 - \partial_\mu \bar{F}_2 \partial_\nu \bar{A}_1 )\big] \,.
\end{split}
\end{equation}

\end{widetext}

In~(\ref{chiral-1-2}) $\partial_\mu \equiv \partial / \partial y^\mu$,
while in~(\ref{antichiral-1-2}) $\partial_\mu \equiv \partial / \partial \bar{y}^\mu$.
%

Thus the star product 
of chiral fields is chiral, and the star product of antichiral fields is antichiral.  One may again note that the reality condition is satisfied,
\be \label{reality-1-2}
 \overline{(\Phi_1 * \Phi_2)} = \bar{\Phi}_2 * \bar{\Phi}_1  \ .
\ee


\subsection{Non-associativity and Weyl ordering}


As usual,
\begin{align} \label{order-1-2}
\Phi_1 * \Phi_2 &\ne \Phi_2 * \Phi_1 \\
 \bar{\Phi}_1 * \bar{\Phi}_2 &\ne \bar\Phi_2 *\bar \Phi_1
\end{align}
%
but here the difference persists even if one isolates (say) the $\theta\theta$ terms and integrates over space.  

When constructing a Lagrangian this would lead to different theories,
depending on the ordering of the superfields.  Following~\cite{Seiberg}, for
example, the Lagrangian can be specified by requiring
products of superfields to be Weyl ordered.  Then a Lagrangian will
get no extra contributions from noncommutativity from terms quadratic
in chiral or in antichiral fields, because the terms proportional to
$\tta\tta$ or $\btta\btta$ that involve $C$ or $\bar{C}$ are
antisymmetric under interchange of the two superfields.

The situation is more complicated for three or more fields, because the
star product~(\ref{star}) is not associative,
\be \label{phi-1-2-3}
\Phi_1 * (\Phi_2 * \Phi_3) \ne (\Phi_1 * \Phi_2) * \Phi_3 \ .
\ee
This is a consequence of having both $Q$ and $\bar{Q}$ in the star
product~(\ref{star}), with $\{ Q, \bar{Q} \} \ne 0$.  For discussion
of associativity of star products see for example~\cite{Klemm}.

We deal with this by defining for a non-associative product a natural Weyl ordering given by
\begin{equation} \label{Weyl-1}
\begin{split}
{\textrm{\bf W}}(f_1 (f_2 f_3)) & \equiv \frac{1}{6}\big[ f_1(f_2f_3)+f_2(f_1f_3) + f_2(f_3f_1)
\\  & \quad +
				                     f_1(f_3f_2)+f_3(f_1f_2) + f_3(f_2f_1) \big]\\
&= \frac{1}{6}\big[ f_1(f_2f_3+f_3f_2) + f_2(f_1f_3+f_3f_1) 
\\  & \quad + f_3(f_1f_2 + f_2f_1) \big] \,.
\end{split}
\end{equation}
and similarly  for ${\textrm{\bf W}}((f_1 f_2) f_3)$.
One can follow this by Weyl ordering the result in the normal way and find that
\be 
{\textrm{\bf W}} \left[{\textrm{\bf W}} (f_1 (f_2 f_3))\right]=
{\textrm{\bf W}} \left[{\textrm{\bf W}} ((f_1 f_2) f_3)\right] 
\equiv  {\textrm{\bf w}}  (f_1 f_2 f_3) \,.   \label{WW}
\ee
%
%
It should be clear that for the star product of just two superfields,
the second Weyl ordering leaves the result unchanged.  We use the
double Weyl ordering just described to unambiguously define any
Lagrangian in the noncommutative space given
by~(\ref{xtta})-(\ref{xx}). As an example, we will write down the
Wess-Zumino Lagrangian in noncommutative Minkowski superspace.  

\subsection{The Lagrangian}

It is useful to record some steps in the calculation of the product of three chiral fields.  Since the star product of two chiral
fields is chiral, from~(\ref{chiral-1-2}) we can obtain the $A_{12}$, $\psi_{12\gamma}$, and $F_{12}$ components
of the chiral field $\Phi_{12} = \Phi_1 * \Phi_2$ as
\begin{equation}
\begin{split}
A_{12} &= A_1A_2 - C^{\al\bt}\psi_{1\al}\psi_{2\bt} -{\textrm{det}}C F_1F_2  \\
\psi_{12\gamma} &= (A_1\psi_{2\gamma} + A_2\psi_{1\gamma}) +
\cab\big[\epsilon_{\bt\gamma}(\psi_{1\al}F_2 - \psi_{2\al}F_1) \\ 
 & \quad + \bcab \smaa\sigma_{\gamma\dot\bt}^\nu(\partial_\mu A_1 \partial_\nu \psi_{2\bt} 
					 - \partial_\mu A_2 \partial_\nu \psi_{1\bt} ) \big] \\
F_{12} &=(F_1A_2+A_1F_2 - \psi_1\psi_2) +
 2\bar{C}^{\mu\nu}\partial_\mu A_1 \partial_\nu A_2  \\
& \quad 
	+\cab\bcab \smaa\sigma_{\bt\dot\bt}^\nu(\partial_\mu A_1 \partial_\nu F_2 
					 - \partial_\mu A_2 \partial_\nu F_1 )
\end{split}
\end{equation}
Then, the star product of three chiral fields is
\begin{widetext}  \vglue -18 pt
%
\begin{equation} \label{chiral-12-3}
\begin{split}
\left( \Phi_1(y,\tta)\, * \,\Phi_2(y,\tta) \right)\, 
		* \,\Phi_3(y,\tta)&= 
A_{12}A_3 - C^{\al\bt}\psi_{12\al}\psi_{3\bt} -{\textrm{det}}C F_{12}F_3 
+ \sqrt{2}\tta^\gamma\Big( A_{12}\psi_{3\gamma} + A_3\psi_{12\gamma} \\
	& \quad +
    \cab\big[\epsilon_{\bt\gamma}(\psi_{12\al}F_3 - \psi_{3\al}F_{12})  
    	+\bcab \smaa\sigma_{\gamma\dot\bt}^\nu(\partial_\mu A_{12} \partial_\nu \psi_{3\bt} 
					 - \partial_\mu A_3 \partial_\nu \psi_{12\bt} ) \big]\Big)\\
 & \quad +\tta\tta\big[ F_{12}A_3+A_{12}F_3 - \psi_{12}\psi_3 +
    2\bar{C}^{\mu\nu}\partial_\mu A_{12} \partial_\nu A_3 \\
&\quad  +\cab\bcab \smaa\sigma_{\bt\dot\bt}^\nu(\partial_\mu A_{12} \partial_\nu F_3 
					 - \partial_\mu A_3 \partial_\nu F_{12} )\big] \,,
\end{split}
\end{equation}   \vglue -5 pt
\end{widetext}  \vglue -25 pt
From~(\ref{chiral-12-3}), the only $C$-dependent term that will contribute to the Wess-Zumino Lagrangian from the double Weyl ordered product 
$ {\textrm{\bf w}} (\Phi_1(y,\tta)\, * \,\Phi_2(y,\tta)\, * \,\Phi_3(y,\tta))$
comes from the $A_{12}F_3$ term.  The contribution from this term is proportional to $-{\textrm{det}}CF_1F_2F_3$, which is Lorentz invariant.
For the star product of three antichiral fields,
one finds a contribution proportional to $-{\textrm{det}}\bar{C}\bar{F}_1\bar{F}_2\bar{F}_3$.

There is no extra contribution to the Wess-Zumino Lagrangian coming
from the kinetic energy term.  From $\bar{\Phi} * \Phi$ there is a term $S^{\mu\nu} \partial_\mu \bar{F}\partial_\nu F$ from the star product, where 
$S^{\mu\nu} \equiv \cab\bcab\smaa\snbb$ is symmetric.  However, it is precisely cancelled when one adds $\Phi * \bar\Phi$ in doing the Weyl ordering.

We find the following simple result for the Wess-Zumino Lagrangian with one chiral $\Phi$ and one  antichiral field $\bar\Phi$,
%
\begin{equation} \label{W-Z}
\begin{split}
\mathcal{L}&={\textrm{\bf w}}\Bigg[\int d^2\tta\tta \, 
			d^2 \btta\btta \,\bar{\Phi}*\Phi
+\int d^2\tta\,\bigg(\frac{1}{2}m\Phi * \Phi 
+ \frac{1}{3}g\Phi   \\  &  \qquad  *\Phi * \Phi\bigg)
+\int d^2\btta\,\left(\frac{1}{2}m \bar{\Phi} * \bar{\Phi}
+\frac{1}{3}g \bar{\Phi} * \bar{\Phi} * \bar{\Phi}\right) \Bigg] \\
&=\mathcal{L}(C=0)
\\ & \quad -\frac{1}{3}g{\textrm{det}}C F^3 -\frac{1}{3}g{\textrm{det}}\bar{C}\bar{F}^3
+\textrm{total derivatives}\,.
\end{split}
\end{equation}
%
This Lagrangian is Hermitian and Lorentz invariant.

\eject

\section{Summary} \label{sec:summary}	

Our goal has been to find a theory that works in Minkowski space that explores non-anticommutativity of the supercoordinates $\theta$ and $\bar\theta$.
We have shown a consistent set of commutation and anticommutation relations for the full set of coordinates $x$, $\theta$, and $\bar\theta$ (or equivalently $y$ or $\bar y$, $\theta$, and $\bar\theta$).
We have found a star product that reproduces all the coordinate commutation relations, and use this star product to define multiplication of arbitrary functions.

The star product is real, meaning it maintains the standard relations obeyed by involutions of products of functions.  This in turn means products that are Hermitian with no star-multiplication are also Hermitian with star-multiplication, after Weyl ordering.  Any Lagrangian extended to noncommutative space using star-products and Weyl ordering will necessarily remain Hermitian.  Further, the star-product maintains the chirality of products of chiral fields, and the antichirality of products of antichiral fields.

The star-product in this work is not associative, in keeping with a general theorem of Klemm, Penati, and Tomassia~\cite{Klemm}.  However, this interesting feature causes little trouble after making a natural modification of the Weyl ordering procedure.  Also, the basic commutation relation between the components of $\theta$ violates Lorentz invariance.  The example Lagrangian we studied, the super-noncommutative Wess-Zumino model, gained only Lorentz invariant modifications, but this cannot be expected to occur in general.

There are a number of potentially interesting directions to pursue in future work.  One clearly wants to extend the present supercoordinate algebra to gauge theories, and to explore potential phenomenological consequences.  One would also like to study connections to string theory and attempt a derivation of the present commutation relations from a string model.  One may also define an explicit connection between operators in noncommutative space and their commutative space symbols, and derive the star-product from it.  The current star product may be just the expansion to second order in deformation parameter $C$ of one found this way. We should note that if this proves to be the case, the results of the present paper will still hold. To this order the star-product we have is unique in satisfying the requirements of giving the supercoordinate commutation relations and of being real.



 \begin{acknowledgments}
 
We thank Chris Carone and Marc Sher for useful discussions. We also thank the NSF for support under Grant PHY-0245056.  

\end{acknowledgments}

\end{document}